\documentclass[12pt]{article}
\usepackage{amssymb} 
\usepackage{graphicx}
\usepackage{amsmath}

\def\d{\partial}
\def\l{\left(}
\def\r{\right)}

\newcommand{\be}{\begin{equation}}
\newcommand{\ee}{\end{equation}}
\newcommand{\ba}{\begin{align}}
\newcommand{\ea}{\end{align}}
\newcommand{\bg}{\begin{gather}}
\newcommand{\eg}{\end{gather}}
\newcommand{\bseq}{\begin{subequations}}
\newcommand{\eseq}{\end{subequations}}

\makeatletter

\def\lsim{\compoundrel<\over\sim}
\def\compoundrel#1\over#2{\mathpalette\compoundreL{{#1}\over{#2}}}
\def\compoundreL#1#2{\compoundREL#1#2}
\def\compoundREL#1#2\over#3{\mathrel
         {\vcenter{\hbox{$\m@th\buildrel{#1#2}\over{#1#3}$}}}}
\makeatother

\begin{document}
\begin{flushright}
HUTP-06/A0009\\
INR/TH-05-2006
\end{flushright}
\begin{center}
{\large \bf
Spontaneous breaking of Lorentz invariance, black holes and
perpetuum mobile of the 2nd kind}
\\
\medskip
S.L.~Dubovsky$^{a,b}$, S.M.~Sibiryakov$^{b}$
\medskip
\\
$^a${\small Jefferson Physical Laboratory,
Harvard University, Cambridge, MA 02138, USA}\\
$^b$ {\small Institute for Nuclear Research of the Russian Academy of
Sciences,\\  
60th October Anniversary prospect, 7a, 117312 Moscow, Russia
}
\end{center}

\begin{abstract}
We study the effect of spontaneous breaking of Lorentz
invariance on black hole thermodynamics. We consider a scenario
where Lorentz symmetry breaking manifests itself by the difference of
maximal velocities attainable by particles of different species in a
preferred reference frame. The Lorentz breaking sector is represented
by the ghost condensate. We find that the notions of black hole
entropy and temperature loose their universal meaning. In particular,
the standard derivation of the Hawking radiation yields that a black
hole does emit thermal radiation in any given particle species, 
but with temperature depending on
the maximal attainable velocity of this species. We demonstrate that
this property implies violation of the second law of thermodynamics,
and hence, allows construction of a perpetuum mobile of the 2nd
kind. We discuss possible interpretation of these results.    
\end{abstract}

\section{Introduction}
It is highly non-trivial that the laws of thermodynamics hold in the 
presence of
gravitational interactions. Indeed, gravitating systems are generically
unstable against collapse resulting in the formation of black holes,
{\it i.e.}, curvature singularities shielded
by event horizons. Classical no-hair theorems
state that black holes are characterized by just a few
parameters (mass, electric charge and angular momentum).  So one may worry
that entropy can be lost behind
the black hole horizons invalidating the second law of
thermodynamics. However, as 
first suggested by Bekenstein
\cite{Bekenstein:1973ur}, it is natural
to assign to black holes the entropy proportional to the horizon area.
With this assignment the net entropy of a black hole and the outer
region never decreases: 
in this way the second law of thermodynamics holds for systems
including black 
holes in general relativity (GR).

The Bekenstein proposal acquires a remarkable
physical justification due to the Hawking effect 
\cite{Hawking:1974sw}. A black hole with
mass $M$  emits thermal radiation with 
temperature $T_H=\l 8\pi M\r^{-1}$ (we set the 
Newton constant equal to one, $G=1$). It is
important that the temperature of the radiation is {\em universal} for
{\em all} species of particles. This allows to consider black hole as
a body with well-defined temperature $T_H$ and the Bekenstein entropy $S_B$
is related to the black hole energy (mass) in the usual way, 
\[
dM=T_HdS_B\;.
\]  
These properties of black holes in GR are believed to reflect 
the fundamental principles of quantum theory.
In particular,
the second law of 
thermodynamics follows from unitarity of quantum physics 
(see e.g.~\cite{weinberg}).
The validity of thermodynamical description of black holes in GR 
is consistent with a 
possibility of constructing a UV
completion of GR in terms of the microscopic quantum theory with
conventional 
properties, where presumably
the Bekenstein
entropy would be reproduced by counting of the microscopic states
of the black hole. Indeed, this was
achieved~\cite{Strominger:1996sh,Maldacena:1996gb} in string theory for 
certain classes of extremal black holes.

One may wonder whether these properties are specific to GR or persist in its
extensions as well. Stated otherwise, it is worth exploring what principles
of GR are crucial for the validity of thermodynamical description of black
holes. Better understanding of these issues may shed light on the
very basic 
principles of quantum gravity and help to explain why we observe gravity the
way it is. 

In this paper we explore the role of microscopic 
Lorentz invariance.
Recent observation of the cosmic acceleration motivated
attempts to modify gravity in the infrared and a number of consistently
looking effective field theories doing this job were
constructed~\cite{Dvali:2000hr}--\cite{Gorbunov:2005dd}. 
Except for the Dvali--Gabadadze--Porrati model~\cite{Dvali:2000hr} all these
theories introduce spontaneous breaking of Lorentz invariance.
In other words, in all these models a non-trivial tensor (vector) condensate
different from the metric is present in the ground state. 
It is natural to ask how spontaneous
Lorentz symmetry breaking affects black hole physics.

It should be stressed that all the models with IR modification of
gravity mentioned above are lacking UV 
completion so far. In some cases, there are arguments indicating
that UV completion is not possible in terms of conventional field theory or
weakly coupled string theory~\cite{Adams:2006sv}. However, arguments presented
in \cite{Adams:2006sv} directly apply only to theories possessing stable
Lorentz-invariant vacuum, whereas in the proposed theories with
spontaneous Lorentz symmetry breaking the would
be Lorentz-invariant vacuum suffers from the ghost
instability and it is unclear whether it makes sense to include it in the
space of the physical states at all. 
Black hole thermodynamics provides 
an alternative way to probe the UV physics of
Lorentz violating models. 
Indeed, it is known for a long time that  
 black holes are unique IR probes of the microscopic theory. For
 instance, in agreement with
string theory, black hole physics implies 
that global symmetries cannot be exact in quantum gravity. 
An interesting extension of this statement to the
case of gauge symmetries was suggested recently in
\cite{Arkani-Hamed:2006dz}. 
It is definitely of interest to find more examples where black
hole physics can provide information about UV completion of the effective
theories.  
If it turns out that Lorentz violating models are not
capable of reproducing the success of black hole thermodynamics in GR,
this 
will indicate that there are fundamental difficulties with their
embedding into microscopic theory and may suggest hints on
necessary properties of the would-be UV completion (if it exists).

We consider modification
of the black hole thermodynamics due to spontaneous Lorentz
symmetry breaking. More specifically, we
study what happens if Lorentz symmetry breaking manifests
itself by the difference of maximal velocities
attainable by particles of
different species in a preferred reference frame. 
Note that if one assumes that  Lorentz symmetry is spontaneously
broken in a hidden sector, then, generically, leading operators 
mediating Lorentz symmetry breaking into the visible
sector result precisely in this effect.
This scenario has been extensively studied in 
flat space-time \cite{Colladay:1998fq, Coleman:1998ti}. 
To study dynamics of such models in
nonlinear gravitational backgrounds one needs to specify 
the structure of the Lorentz breaking sector.
Here we consider a setup where hidden sector is described by 
the simplest of the 
Lorentz symmetry breaking
models,
namely by the ghost condensate model of 
Ref.~\cite{Arkani-Hamed:2003uy}.
 
\section{Setup} 
\label{setup}
The hidden sector contains
a single scalar field $\phi$  with the action 
\be
\label{ghact}
S_\phi=\Lambda^4\int\sqrt{-g}P(X)d^4x\;,
\ee
where\footnote{We use the (+,-,-,-) signature of the
  metric.} 
 $X=\d_\mu\phi\d^\mu\phi$, and  
$P(X)$ is a function with the minimum at $X=1$.  
In the action (\ref{ghact}) we dropped  
terms with
more than one derivative acting on $\phi$, which are generically
present in the ghost condensate action. These terms are not important for our
purposes and we will comment on their effects later. 
We assume that the effective 
cosmological constant is fine-tuned to zero, {\it i.e.}, $P(1)=0$.
Then the system (\ref{ghact}) has a family of
solutions with $X=1$ and vanishing energy-momentum tensor. 
We are interested in two solutions from
this family: the vacuum solution, preserving spatial isotropy and
homogeneity, and the black hole solution.
The vacuum solution  is
\bseq
\label{vac}
\begin{align}
\label{gvac}
&g_{\mu\nu}=\eta_{\mu\nu}\;,\\
\label{phivac}
&\phi=t\;.
\end{align}
\eseq
Note, that the shift symmetry of the action (\ref{ghact})
\be
\label{shift}
\phi\to\phi+c
\ee
implies time translation invariance of the vacuum (\ref{vac}).

Let the visible sector be represented by a massless minimally coupled
scalar field $\psi$. 
The shift symmetry (\ref{shift}) and the reflection symmetry
$\phi\to-\phi$ fix the form of the leading operator
mediating Lorentz symmetry breaking to the visible sector. With this
operator included the action for the field $\psi$ takes the form, 
\be
\label{oper}
S_{\psi}=\int\sqrt{-g}\left(\frac{(\d_\mu\psi)^2}{2}+
\frac{\varepsilon (\d_\mu\phi\d^\mu\psi)^2}{2}\right)d^4x\;,
\ee
where $\varepsilon$ is a dimensionless
parameter. We do not need to 
assume that the parameter $\varepsilon$ is small;
rather, we consider $|\varepsilon|\lsim 1$ which is enough for the
validity of the effective action (\ref{oper}) up to the cutoff scale
$\Lambda$.   
It follows from the form of the action (\ref{oper}) that
the field $\psi$ propagates in the effective 
metric\footnote{We take into account 
that for configurations with $X=1$ there is a proportionality between 
$\det g_{\mu\nu}$ and $\det\tilde g_{\mu\nu}$ with
constant coefficient,
$\det g_{\mu\nu}=(1+\varepsilon)\det \tilde g_{\mu\nu}$.}
\be
\label{geffec}
\tilde g^{\mu\nu}=g^{\mu\nu}+\varepsilon \d^\mu\phi \d^\nu\phi\;.
\ee 
Consequently, the propagation velocity of the field $\psi$ in the
vacuum (\ref{vac}) is equal to
\be
\label{vvac}
v=\frac{1}{\sqrt{1+\varepsilon}}\;.
\ee
Positive (negative) values of 
$\varepsilon$ correspond to
subluminal (superluminal) propagation of the field $\psi$.

Solution describing a black hole of mass $M$ in the
coordinate system regular at the horizon
has the form \cite{Mukohyama:2005rw}
\bseq
\label{bh}
\begin{align}
\label{gbh}
&ds^2=d\tau^2-\frac{2M dR^2}{r(\tau,R;M)}
-r^2(\tau,R;M)d\Omega^2\;,\\
\label{phibh}
&\phi=\tau\;,
\end{align}
\eseq
where
\be
\label{abh}
r(\tau,R;M)=\l\frac{3}{2}\sqrt{2M}\l R-\tau\r\r^{2/3}
\ee
The metric (\ref{gbh}) is nothing but the ordinary Schwartzschild metric
with the mass $M$ in the reference frame of free falling observers 
(Lemaitre reference frame). 
Again, the shift symmetry (\ref{shift})
implies that this solution is stationary. Note that the $\phi$-field 
configuration is smooth
for  this solution so that the effective theory (\ref{ghact}) breaks down
only in the vicinity of the black hole singularity.

The effective metric felt by the field $\psi$ in the black
hole background (\ref{bh}) has the form
\be
\label{tildegbh}
d\tilde s^2=\frac{d\tau^2}{(1+\varepsilon)}-\frac{2MdR^2}{r(\tau,R;M)}
-r^2(\tau,R;M)d\Omega^2\;.
\ee
The coefficient in front of $d\tau^2$ can be absorbed by the rescaling
\be
\label{tildetau}
\tau\mapsto\tilde\tau=\frac{\tau}{\sqrt{1+\varepsilon}}\;.
\ee 
The subsequent rescaling 
\be
\label{tildem}
M\mapsto\tilde M=(1+\varepsilon)M~,~~~~
R\mapsto\tilde R=\frac{R}{\sqrt{1+\varepsilon}}
\ee
casts the effective metric (\ref{tildegbh}) 
into the form (\ref{gbh}) with $\tau$, $R$ and $M$
replaced by $\tilde \tau$, $\tilde R$ and $\tilde M$. 
One observes that 
the metric felt by the field $\psi$
in the
coordinates $\tilde\tau, \tilde R$ 
is again the Lemaitre metric, but now corresponding to the black hole
with the mass $\tilde M$.
In particular, the effective metric has a horizon
at 
$r(\tilde\tau,\tilde R;\tilde M)\equiv r(\tau,R;M)=2\tilde{M}$. 
Consequently, as one could have expected, the black hole horizon
appears larger for subluminal particles and smaller for superluminal.
This is a clear signal that black hole thermodynamics in this
setup is crucially different from the conventional case. Indeed, the 
horizon area is not a universal notion any longer --- different
particle species see different black hole horizons and it is unclear 
what entropy should be assigned to the black hole.

Moreover, the coincidence of the effective metric in the coordinates 
$\tilde\tau, \tilde R$ with  
the metric of a black hole with
the mass $\tilde M$ enables one, at least naively, to carry out
the  
common derivation  
(see e.g. \cite{Hawking:1974sw,Unruh:1976db,Gibbons:1976ue,Parikh:1999mf,Berezin:1999nn}) 
of the Hawking radiation.
One obtains that
the black hole 
emits thermal radiation of $\psi$-particles 
characterized in the tilded coordinate frame by effective 
temperature $\tilde T_\psi=(8\pi \tilde M)^{-1}
=(1+\varepsilon)^{-1} T_H$, where
 $T_H=(8\pi M)^{-1}$ is
the usual Hawking
temperature of the black hole with the mass $M$.
The effective temperature $\tilde T_{\psi}$ is defined with
respect to the rescaled time $\tilde\tau$. Rescaling back to the
physical time $\tau$ we obtain the physical temperature of the emitted
$\psi$-radiation 
\be
\label{temp}
T_\psi=\frac{T_H}{(1+\varepsilon)^{3/2}}=v^3T_H\;.
\ee  
This result implies that the temperature of the Hawking 
radiation emitted by the
black hole in particles of a given type
depends on the (maximal attainable) velocity of the propagation
of these particles. Therefore, 
with Lorentz invariance being spontaneously broken 
the notion of a black hole temperature becomes ill-defined, as the
temperature depends
on the type of particles used to measure it. This is an alarming
property, and, indeed, we are going to show that it leads to the violation
of the second law of thermodynamics and thus opens up a possibility to
construct a perpetuum mobile of the second kind. Later, we will discuss some 
possible loopholes in the naive derivation of the Hawking radiation
suggested above.

\section{Perpetuum mobile of the 2nd kind: construction manual}  
We are going to
present a counterexample to the following formulation of the second law of
thermodynamics: {\em a process, whose only result is the transfer of energy
from a cold body to a hot body, is impossible}. In the setup of
Sec.\ref{setup},  
let us consider two types of
particles, $\psi_1$ and $\psi_2$ with different speeds, $v_2>v_1$. Let us take
a black hole and surround it by two shells, $A$, $B$. We assume that the
shell $A$ interacts only with the field $\psi_1$, while the shell $B$ --- only
with the field $\psi_2$.  We choose the temperatures of the shells to satisfy
\be
\label{inequ}
T_2>T_B>T_A>T_1\;,
\ee
where
$T_1$, $T_2$ are the temperatures of the Hawking radiation of the
black hole in particles $\psi_1$, $\psi_2$, respectively. Now, since
$T_A>T_1$ there will be a net flux of energy $F_1(T_A,T_1)>0$ 
from the shell $A$ into
the black hole carried by the particles $\psi_1$. 
On the other hand, as $T_B<T_2$, the net flux $F_2(T_B,T_2)$ of energy
from the shell $B$ into the black hole, carried by the particles 
$\psi_2$, is negative, $F_2(T_B,T_2)<0$. 
In the conventional case without Lorentz violation the functions 
$F_i$, $i=1,2$, are given by the Steffan-Boltzman formula,
\be  
\label{fT}
F_i(T,T')=\frac{\pi^3}{15}(2M)^2\bigl(\Gamma_i(T) T^4-
\Gamma_i(T') T'^4\bigr)\;, 
\ee
where $\Gamma_i(T)$ are the grey body factors which depend on the
spin of the particles and slowly vary with temperature.
We do not need the explicit form of the functions $F_i$ in the case of
Lorentz violation.
It is sufficient for our argument 
that $F_i$ fulfill the following requirements:\\
$F_i(T,T')=0$ at $T=T'$,\\
$F_i(T,T')$ grow with $T$ at fixed $T'$.\\
Then 
one can choose the temperatures of the shells in such a way that the
two energy fluxes compensate each other, 
$F_1(T_A,T_1)+F_2(T_B,T_2)=0$,
and the black hole mass
stays constant. 
This can be satisfied simultaneously with the inequalities
(\ref{inequ}).
So, for an outer observer the state of the black hole does not
change, 
and the only result of the process
under consideration is the transfer of energy from the shell $A$ to
the shell $B$. As $T_A<T_B$, 
the second law of thermodynamics is violated by this process. In other
words, an outer observer is forced to conclude that the entropy of the
system decreases.

\section{Discussion} 
Definitely, the above conclusion is puzzling,
so at this point one may wonder whether our derivation is too
superficial. At any rate, it is worth asking what
conclusion is to be drawn from our observation. 
It is instructive to divide the 
potential explanations of the strange behavior found here 
into the following three classes
(not necessarily excluding each other).
\newline
{\it (i)} The presented description of the Hawking radiation in the ghost 
condensate is correct, but there is some subtle way in which a low energy
effective theory forces our perpetuum mobile to change its state 
so that the entropy actually increases.
\newline
{\it (ii)} The derivation of the Hawking radiation using only low
energy theory is incorrect.
\newline
{\it (iii)} The presented description of the Hawking radiation in the ghost 
condensate is correct, and the violation of the second law of
thermodynamics within a low energy effective theory is a physical effect. 
According to the discussion in Introduction this means that 
the UV completion of the ghost condensate, if it exists at all, has very
unusual properties.

Let us start with discussing the possibility {\it (i)}.  First, a
possible objection to the scheme of the perpetuum mobile presented
above could be that it does not take into account the Hawking
radiation of gravitons. One way to get around this objection is to
introduce a mildly large number $N\sim |\varepsilon|^{-1}$ 
of fields $\psi_1$ and $\psi_2$
so that the entropy transferred between the shells is much larger than
the entropy radiated in gravitons. 
In this way the effect of gravitons
is made negligible. Note that the needed number of
species $N$ is independent of the mass of the black hole and is
determined solely by the parameter $\varepsilon$.
Another way to get around the above objection is to include gravitons
into consideration and make them play the role of one of the fields, say
$\psi_2$ (then, the field $\psi_1$ must be subluminal). The shell $A$
can be made sufficiently thin to interact weakly with gravitons, while
the shell $B$, on the contrary, can be made sufficiently massive to
absorb all the gravitons emitted by the black 
hole.\footnote{It should be
noted, though, that it is not obvious that a shell $B$ with the needed
properties can indeed be constructed (c.f. "gravity is the weakest
force" conjecture~\cite{Arkani-Hamed:2006dz}).} So, 
we do not find it plausible  that taking into account  gravitons allows
to avoid the conclusion
that the second law of thermodynamics is violated.

For simplicity, let us assume in the rest of the discussion that only
the field
$\psi_1$ has a non-zero value of $\varepsilon$. Note that during the
work of our 
perpetuum mobile there is a non-zero flux of $\psi_1$-particles into
the black  
hole.
One may argue that because of 
the  direct coupling between $\psi_1$ and the ghost condensate, the ghost 
condensate profile outside the black hole ``remembers'' about the
amount 
of $\psi_1$ particles swallowed by the black hole. In other words, there can
be extra entropy available  for the outside observer which is contained
in the perturbations of the ghost condensate. To see that this is not
the case, 
recall that there is a convenient fluid analogy \cite{Arkani-Hamed:2005gu}
for the ghost condensate. Namely, one introduces a unit four-vector 
\be
\label{u}
u_\mu=\frac{\d_\mu\phi}{\sqrt X}\;.
\ee 
Then the field equations
of the ghost condensate  coincide with the hydrodynamical equations
describing dynamics of the irrotational relativistic fluid consisting
of two components which cannot mix with each other. The fluid four-velocity
is given by $u_\mu$, $P(X)$ plays the role of the pressure, while $(2XP'-P)$
is the energy density. 
One component has $X\geq 1$
and positive energy density, while the other has $X<1$ and negative energy 
density. For instance, the field equation of 
the ghost condensate
\be
\label{gheom}
\nabla_\mu(P'\nabla^\mu\phi)=0\;
\ee
is a conservation law for the fluid current.
This equation changes in the presence of the field $\psi_1$ coupled to the
effective metric  (\ref{geffec}). One obtains
\be
\label{gheom1}
\nabla_\mu(P' \nabla^\mu\phi)=-\frac{\varepsilon}{2\Lambda^4}
\nabla_\mu(\nabla^\mu\psi_1\nabla_\nu\psi_1 \nabla^\nu\phi)\;.
\ee 
Consequently, $\psi_1$-field plays the 
role of the source for the ghost current.
At moderate $\varepsilon$, $(1+\varepsilon)\sim 1$, 
the
expectation value $<\nabla^\mu\psi_1\nabla_\nu\psi_1>$ can be
estimated in the coordinate system corresponding to the metric (\ref{gbh}) as 
\be
|<\nabla^\mu\psi_1\nabla_\nu\psi_1>|\sim T_{\psi_1}^4\;.
\ee
Thus, the source in (\ref{gheom1}) 
can be considered as a small perturbation as long as
\be
\label{smallTpsi}
\frac{|\varepsilon| T^4_{\psi_1}}{\Lambda^4}\ll 1\;,
\ee  
which is true for a large enough black hole.
Note that unperturbed solution (\ref{phibh}) corresponds to a spherically 
symmetric flow of zero energy fluid into the black hole.
Any small perturbation of the ghost condensate fluid
induced by the $\psi_1$ particles will be carried into the black hole
by the background flow so that at the end
there is no memory left outside about the number of $\psi_1$ particles
emitted by the black hole.

Another possible objection to our construction may be that the 
ghost condensate 
action (\ref{ghact}) is just a low energy effective action and
in general it also contains terms with more derivatives acting on $\phi$.
One effect of these terms is that the ghost condensate exhibits 
Jeans instability \cite{Arkani-Hamed:2005gu}. So, the device 
described above is not,
strictly speaking, a perpetuum mobile. However, the characteristic time
of the growth of the instability is 
\be
\label{Jeenst}
t_J\sim\frac{1}{\Lambda^3}\;,
\ee 
and can be very large\footnote{It is longer than the lifetime of the
  Universe 
if $\Lambda\lsim 10$ MeV.}. The requirement that the
device can operate only for the period of time shorter than $t_J$
together with the inequality (\ref{smallTpsi}) places a rather
mild upper bound on
the amount of entropy and energy  which can be transferred between the
shells. For instance, an estimate for the latter is given by 
(c.f. Eq.~(\ref{fT}))
\be
\label{amount}
E\sim |\varepsilon| \bigl(2\tilde M\bigr)^2 T_{\psi_1}^4 t_J\;, 
\ee  
Taking into account the constraint 
(\ref{smallTpsi}) one obtains
\be
\label{amountupper}
E<E_{max}=\frac{\sqrt{|\varepsilon|}}{\Lambda}\;.
\ee 
This bound is not restrictive, for $\varepsilon=0.01$,
$\Lambda=10$ MeV we obtain $E_{max}=10^{36}$erg which is a huge amount of
energy.

Another effect of the higher derivative terms is that they modify
the black hole solution (\ref{bh}) and lead to the accretion of ghost
condensate
energy into black hole. However, as shown in \cite{Mukohyama:2005rw},
the accretion rate is very slow. The change in the black hole mass
over the time $t_J$ is negligible for black holes heavier than the
Planck mass. So it appears that higher derivative terms generically 
do not prevent
violation of the second law of thermodynamics as well.

The above arguments do not completely exclude the possibility {\it
(i)} and further study is needed. However, to our opinion, this
possibility is unlikely, as the very fact that a notion of horizon
does not have a universal meaning strongly suggests the breakdown of the
standard black hole thermodynamics. Actually, this is already
suggested by the observation that perturbations in the ghost condensate
can carry negative energy and thus violate the null energy
condition. The latter is known to be related to the entropy bounds
\cite{Bousso:2002ju}, and consequently, to the black hole
thermodynamics. So let us turn to the other two possible options
mentioned above.

We proceed to the option {\it (ii)}. There are
several derivations of the Hawking effect in GR. Let us 
discuss 
subtleties which potentially may affect these approaches in our setup. 
First, recall that the original derivation
due to Hawking \cite{Hawking:1974sw} applies to the collapsing body,
while we considered an eternal black hole. Fluid picture makes it likely that
in our case the metric of the collapsing black hole is not  different 
from the usual case. However, a ghost field singularity (caustic)
is likely to be present at the origin even before the horizon forms (see
\cite{Arkani-Hamed:2003uy,KRRZ} for a discussion of these caustics).
At the caustic the gradients of the ghost field are discontinuous, and
the effective metric (\ref{geffec}) is ill-defined.
The discontinuity of the gradients is supposed to be smoothed out by
the higher derivative terms in the ghost condensate action.
So,
strictly speaking, the presence of these caustics makes it impossible
to apply the Hawking derivation for a field $\psi$ having a non-vanishing
coupling to the ghost condensate field
without knowing UV completion of the
ghost condensate theory. On the other hand, it has been argued 
that the caustics
are resolved in an extension of the ghost condensate model 
dubbed gauged ghost condensate 
\cite{Arkani-Hamed:2005gu,Cheng:2006us}. It would 
be interesting to see how our analysis is
modified in the case of the gauged ghost condensate.

Another well-known subtlety of the 
Hawking derivation is the so called trans-Planckian problem 
(see, e.g., \cite{Unruh:2004zk} 
for a recent discussion). 
Indeed, the 
presence of the Hawking radiation results from
the Bogolyubov transformation between the modes of the {\em in-} and 
{\em out-}vacua. 
But the outgoing wave of the 
Hawking radiation gets infinitely blue-shifted in the region near the 
black
hole horizon,
so this calculation implicitly assumes that
one knows the form of the vacuum modes up to the 
trans-Planckian frequencies. 
This assumption is also present in the approach of
Ref.~\cite{Unruh:1976db}, which makes use of the 
eternal black hole metric and thus avoids the subtlety with the 
ghost condensate
caustics. 
The above assumption is justified in the standard
case because the vacuum modes with high frequencies are related to the
soft modes by Lorentz boosts.
However, in the presence of the ghost condensate one has a preferred
reference 
frame (rest frame of the ghost condensate), and the statement that a wave has
large frequency in this frame has an invariant physical meaning. 
Strictly speaking the effective action (\ref{ghact}) applies only to modes 
with frequencies smaller than $\Lambda$ in this reference frame, so again
we conclude that the Hawking radiation may be sensitive to the details
of the UV completion in our setup.
If true, this conclusion is rather interesting because Hawking radiation 
was shown to be independent of the details of UV completion under rather
general assumptions, even allowing for Lorentz invariance breaking in the
UV (while preserving it in the IR)~\cite{Unruh:2004zk}. 

A subtlety of the approach \cite{Gibbons:1976ue} based on the
Eucledian continuation of the black hole metric is that in our setup
the analytic continuation is to be performed differently for different
species, because they feel different effective metrics. Moreover, an
explicit time-dependence of the ghost condensate background makes it
somewhat unclear how to perform Eucledian continuation of the full
theory. However, at the quadratic order where the interaction of the
field $\psi$ with the ghost condensate is described by the effective
metric (\ref{geffec}) there are no apparent obstructions to the
analytic continuation of the effective metric, and the thermal circle
is present in the Eucledian time. This reasoning would be incorrect
if higher order terms describing interaction with the ghost condensate
in the action for the field $\psi$ could not be neglected; 
this would again mean that the Hawking
radiation in our setup is sensitive to the UV physics. Another
consequence of this option would be that the existence of the
thermal Eucledian circle in the low energy action does not imply
thermal behavior in Minkowski signature.  Recently, an example of a
completely different system where this happens was discussed in
\cite{seiberg} with a possible conclusion that such a behavior is an
indication of non-locality in time.

We see that the latter possibility is closely related to the option
{\it (iii)}, namely that a UV completed theory should be very
unusual. As a concrete scenario of a microscopic theory where the
conflict with the second law of thermodynamics is resolved one can
imagine a theory 
containing an infinite tower of fields $\psi_i$ with indefinitely
growing (maximal) 
propagation velocities. 
Such a theory would not possess any black hole
horizons at all --- one would be capable of probing the entire 
black hole
interior using fields with higher and higher velocities.  Then one
would be able to see that the particle content of the "black hole"
changes in our process, and calculate the entropy directly without
relying on the Bekenstein formula, thus avoiding a conflict with the
second law of thermodynamics.

It would be interesting to understand which of the options {\it
(i)--(iii)} is actually realized. 

To conclude, it is worth stressing
that we concentrated on the scenario where the Lorentz violating
sector is represented by the ghost condensate just for the sake of
simplicity. We expect our conclusion that  
the 
standard black hole
thermodynamics breaks down in the presence of spontaneous
Lorentz symmetry breaking to be rather generic and apply to other 
models as well.

{\it Acknowledgments.} 
We thank Nima Arkani-Hamed,  Dmitry Levkov, Shinji Mukohyama, Alberto
Nicolis, 
Valery Rubakov, Mattias Zaldarriaga and especially  
Alexey Boyarsky and Fedor Bezrukov
for helpful discussions. The work of S.S. is supported by RFBR grant
05-02-17363, the grant of the President of the Russian Federation
MK-2205.2005.2 and the grant of the Russian Science Support Foundation.

\end{document}